\documentclass[conference]{IEEEtran}
\usepackage{cite}
\usepackage{threeparttable} 
\usepackage{array}
\usepackage{subfigure}
\usepackage{tablefootnote}
\usepackage{booktabs}
\usepackage{makecell}
\usepackage{stfloats}
\usepackage{amsmath,amssymb,amsfonts}
\usepackage{graphicx}
\usepackage{textcomp}
\usepackage{xcolor}
\usepackage{pifont}
\usepackage{algorithm}  
\usepackage[normalem]{ulem}  
\usepackage{algpseudocode}  
\usepackage{amsmath}  
\usepackage{multirow}
\usepackage{fix-cm}

\usepackage{geometry}

\begin{document}
\date{}

\title{\fontsize{13.8pt}{18pt}\selectfont\bf
Automated C/C++ Program Repair for High-Level Synthesis via Large Language Models
\vspace{-1.4cm}
}

\author{
\IEEEauthorblockN{Kangwei Xu\textsuperscript{1}, Grace Li Zhang\textsuperscript{2}, Xunzhao Yin\textsuperscript{3}, Cheng Zhuo\textsuperscript{3}, Ulf Schlichtmann\textsuperscript{1}, Bing Li\textsuperscript{4}}
\IEEEauthorblockA{\textsuperscript{1}\textit{Chair of Electronic Design Automation, Technical University of Munich (TUM)}, Munich, Germany \\
\textsuperscript{2}\textit{Hardware for Artificial Intelligence Group, Technical University of Darmstadt}, Darmstadt, Germany \\
\textsuperscript{3}\textit{College of Information Science and Electronic  Engineering, Zhejiang University}, Hangzhou, China \\
\textsuperscript{4}\textit{Research Group of Digital Integrated Systems, University of Siegen}, Siegen, Germany \\
Email: $\{$kangwei.xu, ulf.schlichtmann$\}$@tum.de, grace.zhang@tu-darmstadt.de, $\{$xzyin1, czhuo$\}$@zju.edu.cn, bing.li@uni-siegen.de}
\vspace{-1.0cm}
}

\maketitle
\thispagestyle{empty}

\begin{abstract}
In High-Level Synthesis (HLS), converting a regular C/C++ program into its HLS-compatible counterpart (HLS-C) still requires tremendous manual effort. 
Various program scripts have been introduced to automate this process. But the resulting codes usually contain many issues that should be manually repaired by developers. Since Large Language Models (LLMs) have the ability to automate code generation, they can also be used for automated program repair in HLS. However, due to the limited training of LLMs considering hardware and software simultaneously, hallucinations may occur during program repair using LLMs, leading to compilation failures. Besides, using LLMs for iterative repair also incurs a high cost. To address these challenges, we propose an LLM-driven program repair framework that takes regular C/C++ code as input and automatically generates its corresponding HLS-C code for synthesis while minimizing human repair effort. To mitigate the hallucinations in LLMs and enhance the prompt quality, a Retrieval-Augmented Generation (RAG) paradigm is introduced to guide the LLMs toward correct repair. In addition, we use LLMs to create a static bit width optimization program to identify the optimized bit widths for variables. Moreover, LLM-driven HLS optimization strategies are introduced to add/tune pragmas in HLS-C programs for circuit optimization. 
Experimental results demonstrate that the proposed LLM-driven automated framework can achieve much higher repair pass rates in 24 real-world applications compared with the traditional scripts and the direct application of LLMs for program repair.
\end{abstract}

\section{Introduction}
High-Level Synthesis (HLS) tools have achieved significant progress in automatically generating Hardware Description Languages (HDL) such as Verilog from general-purpose programming languages such as C/C++.
This advancement allows software engineers to contribute to hardware design more actively, thereby lowering the expertise barrier in hardware design~\cite{b10,b10.1,b10.2}. However, HLS tools only support a subset of C/C++ programs, necessitating significant manual rewriting to transform regular C/C++ programs into their HLS-compatible C/C++ counterparts (HLS-C)~\cite{b11,b11.1,b12,b13}. 
In other words, given C/C++ programs should be \textit{repaired} before they can be processed by HLS tools to generate the corresponding circuits. 
For example, the use of pointers, recursion, and dynamic memory should be manually rewritten, as such programs are not compatible with the existing HLS tools. In addition, the integer arithmetic in CPU instructions is typically defined with 32/64 bits, whereas in circuit design, the bit widths are customized to enhance hardware efficiency~\cite{b14.2}. 
An unnecessarily large bit width can lead to increased area overhead, high power consumption, and large clock periods in hardware implementation~\cite{b14.3}.

Several methods have been proposed for \textit{program repair} in HLS taking C/C++ programs as input.
 \cite{b11} statically analyzes the pointers in C/C++ programs to reduce global connections. However, the challenge remains when the application accesses memory dynamically with various complex pointers. \cite{b12} provides Domain Specific Languages (DSLs) on top of C++ programs to support recursion in HLS, but control statements need to be manually added. \cite{b13} supports dynamic memory by providing data structure templates. However, only a limited set of data structures is supported, and significant refactoring is still required. \cite{b13.1} presents an HLS backend for generating a customized accelerator using C++ template-based, parameterized types. However, this approach requires the user to manually specify the bit widths. \cite{b14} reduces bit widths through both program profiling and bit analysis. When inputs exceed the typical range, a fallback function is triggered, which is an error-prone process. 

Large Language Models (LLMs) have exhibited great potential in automating both software and hardware design, supporting engineers throughout the design flow from initial concepts, algorithms, and architectures to debugging, verification, and optimization~\cite{b15,b16,b16.1,b16.2,b16.3,b16.4}. Recent studies have demonstrated 
that LLMs can be used to correct syntax and logic errors in Verilog and C/C++ programs \cite{b18,b19,b19.1}. \cite{b16} proposes an iterative and conversational-based approach to generate Verilog codes using LLMs, and \cite{b18} harnesses the capabilities of LLMs as autonomous agents, incorporating human knowledge for reasoning and action planning to automate syntax error fixing for Verilog code. For  C/C++ programs, \cite{b20} introduces iterative approaches that repeatedly query LLMs based on feedback from previous fix attempts to automate program repair. \cite{b21} proposes an LLM-driven program repair approach, where the buggy code is removed, and the LLM directly predicts the correct code given the prefix and suffix context.

Previous methods above use LLMs to generate codes or fix simple errors in existing codes. However, repairing regular C/C++ codes so that they can be converted into HLS-C for hardware design with HLS tools has not yet been well studied. In this paper, we propose an automated LLM-driven C/C++ program repair framework for HLS, which employs an LLM-based search and iterative repair process to eliminate HLS-incompatible errors while enhancing hardware performance. The key contributions of this paper are summarized as follows. 

\begin{itemize}
\item 
We propose an automated C/C++ program repair framework driven by LLMs for HLS with minimal human effort. The proposed workflow covers runtime profiling and C/C++ program repair for hardware generation, verification, and performance optimization.

\item 
To mitigate factual errors (hallucinations) produced by LLMs, a Retrieval-Augmented Generation (RAG) paradigm is introduced to guide the LLMs toward correct repair. An external library containing correct repair templates is built with human guidance and from the manual of HLS tools. These templates are matched through an embedding retrieval mechanism to enhance the quality of input prompts for LLMs, which leads to a 23.33\% increase in the repair pass rate.

\item 
To find the optimal bit widths for variables implemented in hardware, we use the C/C++ code, the corresponding task scenarios, and the description of input data as prompts to create a static bit width optimization program via LLMs. This optimization effectively reduces the bit width and achieves an average of 36.57\%, 33.03\%, and 29.08\% reduction in area, power, and minimum clock period, respectively.

\item 
A joint LLM-script repair mechanism is introduced to pre-repair simple errors at early stages and minimize the cost of using LLMs. With the script providing preliminary repair structures, the repair cost of using LLMs can be reduced by an average of 21.56\%.

\item 
To achieve a better power, performance, and area (PPA) design of the synthesized circuit, successfully repaired HLS-C programs are collected into a potential list for subsequent optimization. Critical code segments with large area, high power, and high latency are identified by HLS tools.  LLMs are used to further optimize these critical code segments to achieve a more efficient PPA design.

\end{itemize}

\begin{figure}[]
\centering	\includegraphics[width=0.95\linewidth]{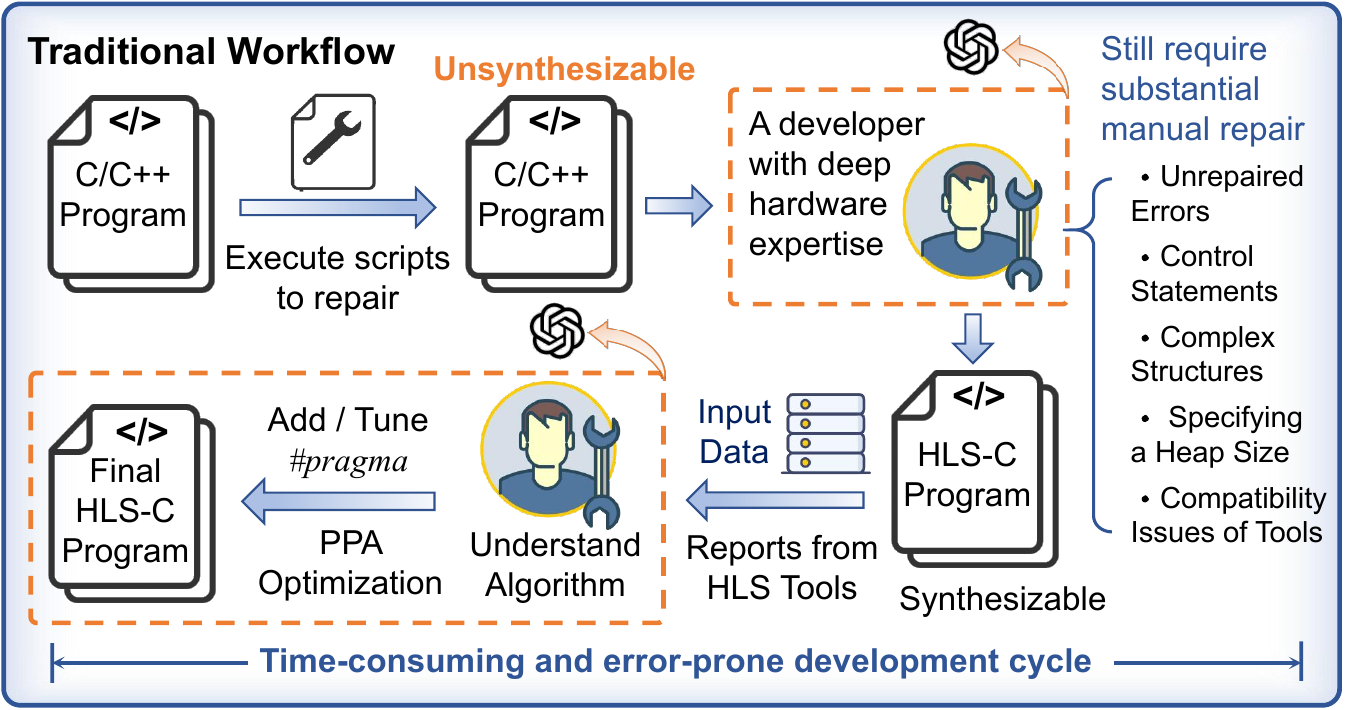}
	\caption{~Traditional workflow for repairing regular C/C++ programs in HLS.}
	\label{fig:tw}
\end{figure}

The rest of this paper is organized as follows. Section~\ref{sec:second} provides the background and motivation of this work. Section~\ref{sec:third} explains the details of the proposed method. The experimental results are provided in Section~\ref{sec:fourth}. Section~\ref{sec:fifth} concludes the paper.

\section{Background and Motivation}\label{sec:second}
Regular C/C++ programs are usually designed to be executed by CPUs; HLS tools often cannot compile them correctly because only a subset of the C/C++ programs, i.e., HLS-compatible C/C++, can be mapped to hardware design directly.

Traditionally, making regular C/C++ programs synthesizable requires developers to possess interdisciplinary expert knowledge in both hardware and software, often involving substantial manual rewriting. It is crucial to eliminate or at least reduce such manual effort to allow designers to focus on logic and performance optimization. Table I summarizes several types of HLS incompatibility types with their corresponding error symptoms and repair edits~\cite{b25}. Key incompatibility types are summarized as follows:

{\setlength{\parskip}{0em} 
$\circ$~\textit{Pointer:} Pointers are strictly forbidden in HLS tools, except for statically analyzable ones. Thus, developers need to manually convert pointer access to array access.

$\circ$~\textit{Dynamic array:} HLS tools do not support dynamic arrays because hardware implementations cannot manage data structures with unbounded size. Hence, dynamic array functions such as \textit{malloc()} and \textit{free()} must be replaced with pre-allocated static arrays.

$\circ$~\textit{Recursion:} This design style requires dynamic storage in a stack of execution states and is not supported by HLS tools. Manual transformation into loops is required.

$\circ$~\textit{Bit Width:} Regular C/C++ programs running on CPUs use standard data types, such as 32-bit for integers, which leads to wasted on-chip resources. Developers need to determine optimal bit widths for variables to achieve a more efficient circuit design. 

\begin{table}[]
\centering	
\caption{~Examples of HLS-incompatible types.}
\includegraphics[width=1.01\linewidth]{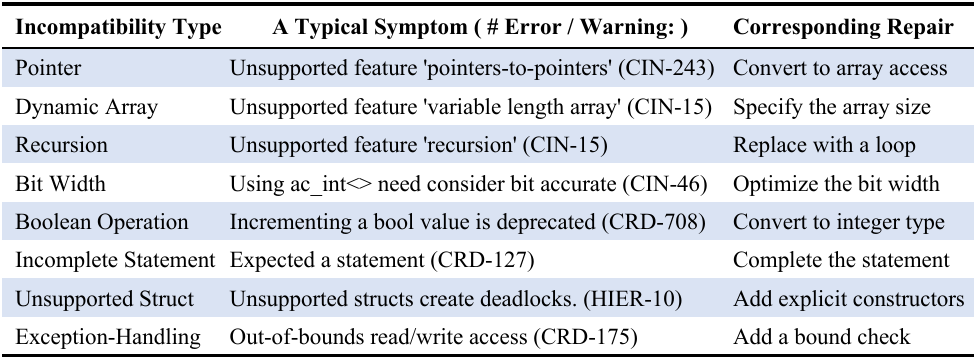}
\label{tab:error}
\vspace{-0.3cm}
\end{table}

$\circ$~\textit{Boolean Operation:} HLS tools do not support assignment operators (+=, $-$=) or increment operators (++, $--$) for Boolean values. An appropriate integer type for these operators needs to be used to replace the Boolean type.

$\circ$~\textit{Incomplete Statement:} HLS tools generate a violation when the case/switch statement does not completely cover the range of values for the object used in conditional expressions. Considering the uncovered values to complete the switch statement is needed.

$\circ$~\textit{Unsupported Structures:} HLS tools do not support virtual functions in C++, necessitating the use of strategy or template methods to achieve polymorphic behavior in hardware designs. 

$\circ$~\textit{Exception-handling:} Violations are reported when there are out-of-bound array reads/writes and illegal shifts. Additionally, whereas CPU programs use stack popping to handle exceptions, on ASIC, exception-handling modules need to be built to deliver the termination message to other modules.}

\begin{figure*}[]
\vspace{-0.2cm}
\centering	\includegraphics[width=1\linewidth]{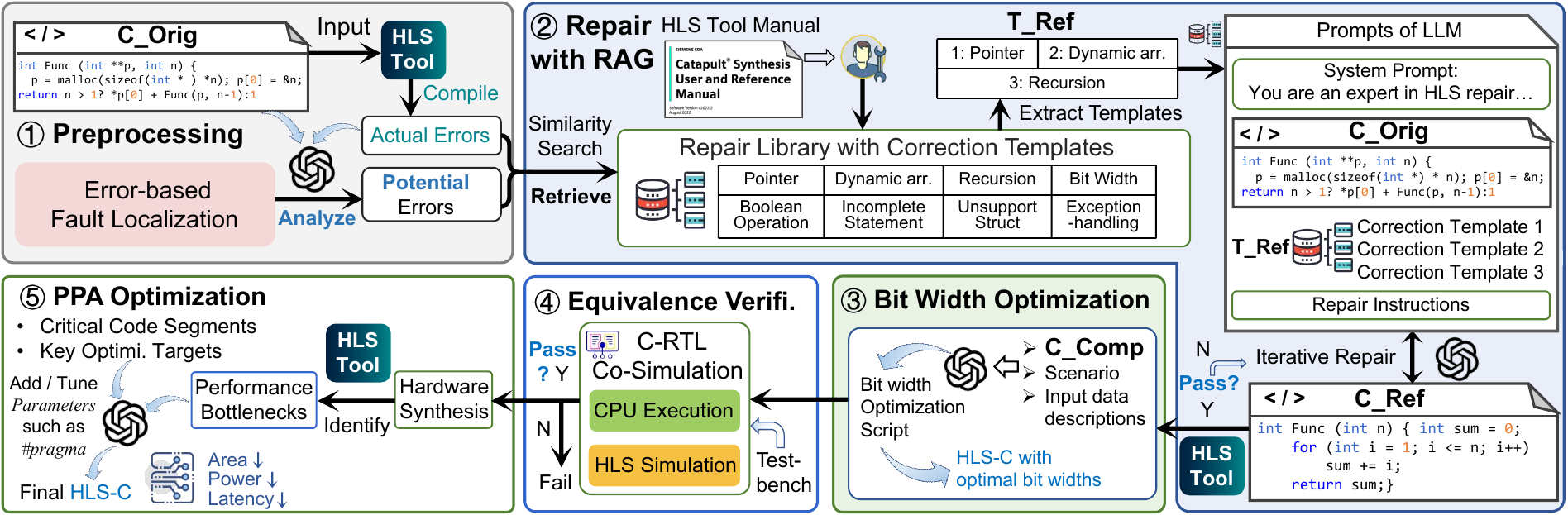}
	\caption{~The proposed LLM-driven automatic C/C++ program repair framework for HLS.}
	\label{fig:workflow}
	\vspace{-0.5cm}
\end{figure*}

Fig.~\ref{fig:tw} illustrates the traditional workflow, where developers rely on scripts to repair the HLS incompatible programs. However, these scripts can only address basic errors, leaving complex or unforeseen issues unresolved. Therefore, substantial manual repair by a developer with deep hardware design expertise becomes essential. After the incompatibility errors are fixed, the developer then applies EDA tools to further optimize the design.

Several techniques have been proposed to address the incompatible issues of regular C/C++ programs in HLS tools \cite{b12,b13,b13.1}. However, these techniques can only partially resolve the aforementioned errors in regular C/C++ programs. Developers still need to continually apply various scripts implementing these individual techniques to enhance the hardware synthesizability of C/C++ programs, thus leading to error-prone and time-consuming development cycles.

To automate program repair, recent studies have reported promising results of directly applying LLMs~\cite{b19,b19.1,b20,b21} to repair C/C++ programs. But this direct application of LLMs for HLS still faces several main challenges: 1) The majority of research on LLM-based program repair focuses on fixing syntax and logical errors~\cite{b18}\cite{b20}, with a notable lack of studies on the automatic structural repair of C/C++ programs for HLS using LLMs; 2) LLMs are prone to producing hallucinations, especially in the interdisciplinary field such as HLS that requires both software and hardware knowledge. It's thus necessary to integrate correct repair guidance into the input prompts of LLMs to improve output response; 3) Using LLMs is costly, and existing approaches have not considered cost-effectiveness in applying LLMs to iteratively repair simple errors.

Contrary to the previous work, the proposed framework is designed to comprehend repair guidelines and combine the relevant design tools/scripts with LLMs to repair C/C++ programs for HLS. It takes regular C/C++ programs as input and automatically generates corresponding HLS-C designs while ensuring correct syntax and logical functionality and minimizing the cost of using LLMs.

\section{LLM-Driven C/C++ Program Repair for High-Level Synthesis}\label{sec:third}

\subsection{Overview of the Proposed Framework} 
As shown in Fig.~\ref{fig:workflow}, the proposed LLM-driven automated C/C++ program framework for HLS consists of five stages: 

\textit{1) Preprocessing:} An original C/C++ program $C\_Orig$ is compiled using the HLS tool, and actual errors are reported. Since the compiler may not be able to detect all errors in a single compilation, the common HLS-incompatible errors, along with the complete program, are fed into LLM to detect other potential errors. An example of the preprocessing stage for detecting and analyzing HLS-incompatible errors is demonstrated in Fig.~\ref{fig:stage1} in Appendix II. After preprocessing, the stage moves to repair with RAG.

\textit{2) Repair with RAG:} A repair library containing correction templates for HLS is manually established, which is built with human guidance and from the official documentation of HLS tools. The Retrieval-Augmented Generation (RAG) technique is employed to search similar repair templates $T\_Ref$ from this library, which can be used to enhance the quality of LLM's prompts while generating a more accurate program $C\_Ref$. If the compilation of $C\_Ref$ fails, the error message will be fed back to the LLM for iterative repair. Otherwise, the successfully compiled program is saved as $C\_Comp$. 

\textit{3) Bit Width Optimization:} To optimize the hardware synthesized by the HLS tool, the existing dataset corresponding to the C/C++ algorithm is used to determine the optimal bit width of variables in $C\_Comp$. Then, the repaired C++ program, corresponding task scenarios, and the descriptions of input data from the dataset are fed into the LLM, which automatically generates a C++ script that is able to assign the optimal bit width by calculating the maximum and minimum values of the variables. 

\begin{figure}
\centering	\includegraphics[width=0.98\linewidth]{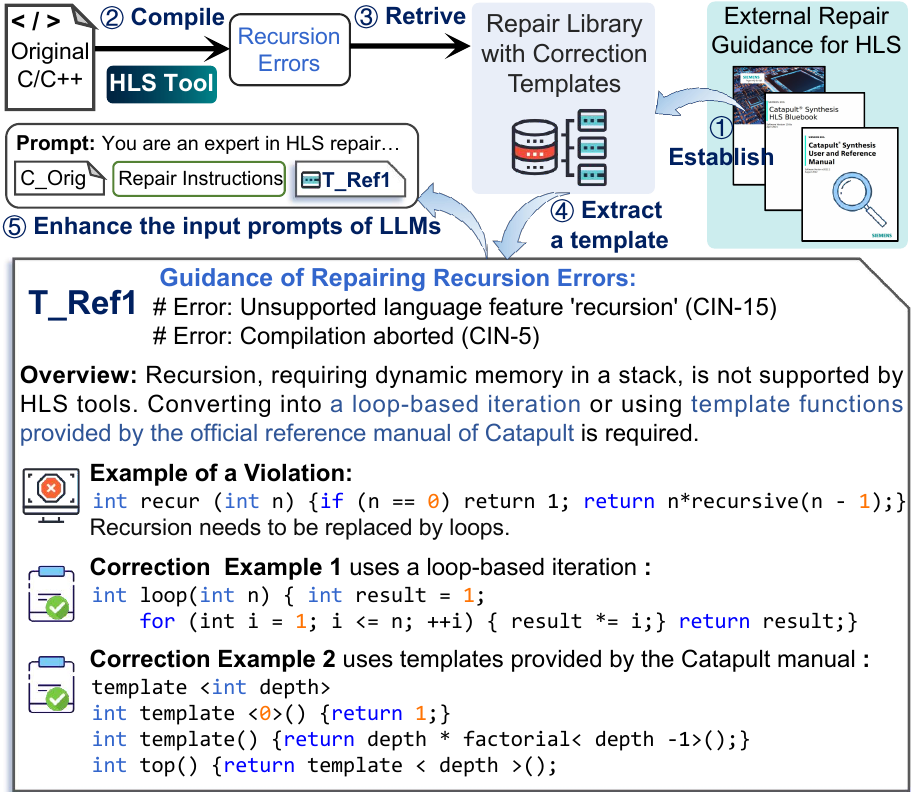}
\vspace{-0.1cm}
	\caption{Example of the LLM using RAG to repair recursion errors.}
	\label{fig:rag}
\end{figure}

\begin{figure}[t]
    \centering
    \begin{subfigure}
        \centering
        \includegraphics[width=1.12\linewidth]{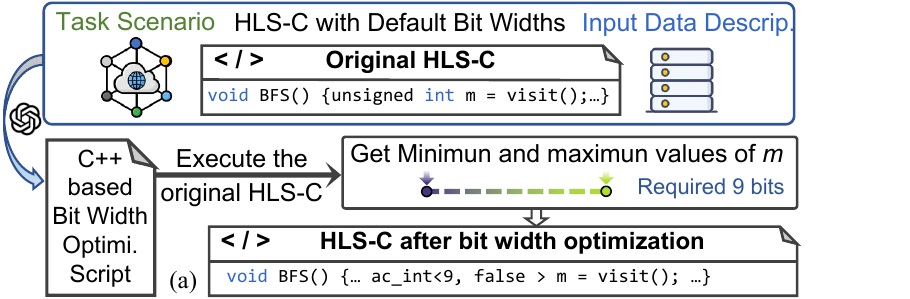} 
    \end{subfigure}
    
    \begin{subfigure}
        \centering
  \includegraphics[width=1\linewidth]{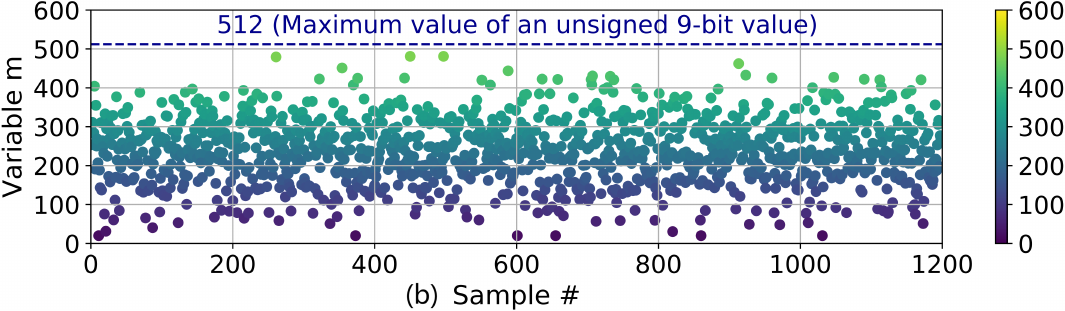}
  \vspace{-0.4cm} 
    \end{subfigure}
    
    \caption{(a) Example of the bit width optimization scheme; (b) 
Distribution of 1200 samples of the variable ‘$m$' from the real-world BFS task. This example illustrates that the C++-based script optimizes the bit width of the variable ‘$m$', which needs only 9-bit instead of the default 32-bit of the \textit{int} type.}
    \label{fig:bop}
\end{figure}

\textit{4) Equivalence Verification:} After synthesizing the HLS-C program into a corresponding RTL code, the C-RTL co-simulation is performed for equivalence verification. Based on the original C++ test benchmarks, the HLS tool automatically compares the RTL simulation results with the $C\_Orig$ simulation results to verify the correctness of the synthesized RTL design.

\textit{5) PPA Optimization of Circuits:} The successfully repaired HLS-C programs are collected into a potential list for further optimization. Critical code segments with large area, high power, and high latency are identified by the HLS tool. The LLM is then used to optimize these key segments by adding/tuning pragmas with the guidance of the HLS official manual. This framework integrates automated interaction between LLM and HLS tools, thereby greatly improving the reliability and efficiency of the repair and optimization process.

An example of repairing a regular C++-based breadth-first search (BFS) algorithm is provided in Appendix I to illustrate the proposed framework. While stages \textit{1)} and \textit{4)} above are straightforward, the key stages \textit{2)}, \textit{3)}, and \textit{5)} above are explained in detail in the following subsections.

\begin{figure}[]
\centering
	\includegraphics[width=1\linewidth]{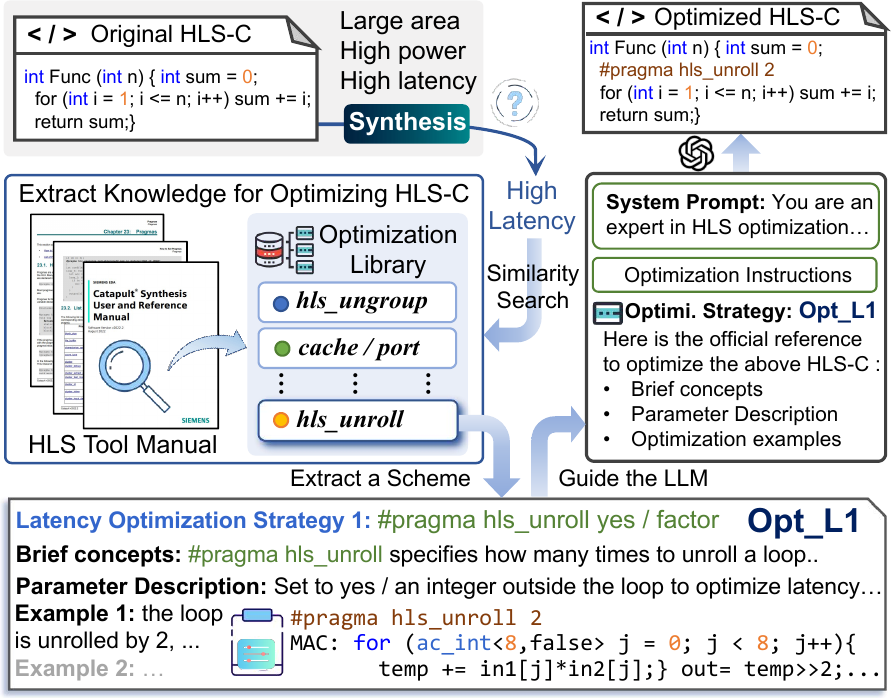}
	\caption{LLM-driven automatic optimization scheme.}
	\label{fig:op}
\end{figure}

\subsection{Retrieval-Augmented Generation (RAG)}
Retrieval-Augmented Generation (RAG) is a technique for enhancing the LLM's capability by incorporating external expert guidance through a retriever. Employing RAG to the LLM for repairing HLS incompatibility errors involves the following stages:
\ding{172} An external repair library containing correction templates is created. Each template integrates the error message from the compiler log, a violation example, corresponding guidance context, and correction examples extracted from the manual of the HLS tool.
\ding{173} After obtaining the compiler error log, a similarity search mechanism, such as a sentence transformer~\cite{b26}, is employed to retrieve the corresponding correction template from the repair library.
\ding{174} Once the correction template with the highest similarity is retrieved, it is extracted and used as part of the prompts provided to the LLM.

Fig. \ref{fig:rag} illustrates the workflow of employing RAG for repairing the HLS-incompatible  ‘\textit{recursion}' error. Initially, when the compiler detects a ‘\textit{recursion}' error, it generates an error log. This log is then used to retrieve the most appropriate correction template in the HLS repair library by the sentence transformer \cite{b26}. Once the correction template with the highest similarity is retrieved, it is used as part of the prompts for the LLM. This correction template ($T\_Ref1$ in Fig. \ref{fig:rag}) includes HLS guidance on how to repair recursion, examples of a violation, and corresponding repaired versions. By integrating this external guidance into the input prompts of the LLM, the output responses are improved significantly. Another example of RAG-based program repair for HLS via the LLM, including the prompts and responses, is demonstrated in Fig.~\ref{fig:stage2} in Appendix II.

\begin{table*}[!t]
\vspace{-0.1cm}
  \centering
  \caption{Comparison of the proposed framework with the traditional scripts and the GPT-4 Turbo Baseline}
  \begin{threeparttable}
    \setlength\tabcolsep{4pt} 
    \resizebox{1\textwidth}{!}{
\begin{tabular}{
  >{\raggedright\arraybackslash}p{1.10cm}
  >{\raggedright\arraybackslash}p{2.88cm}
  >{\centering\arraybackslash}p{0.6cm}
  >{\centering\arraybackslash}p{0.7cm}
  >{\centering\arraybackslash}p{0.7cm}
  >{\centering\arraybackslash}p{0.7cm}
  >{\centering\arraybackslash}p{0.7cm}
  | 
  >{\raggedright\arraybackslash}p{1.10cm}
  >{\raggedright\arraybackslash}p{2.88cm}
  >{\centering\arraybackslash}p{0.6cm}
  >{\centering\arraybackslash}p{0.7cm}
  >{\centering\arraybackslash}p{0.7cm}
  >{\centering\arraybackslash}p{0.7cm}
  >{\centering\arraybackslash}p{0.7cm}
}
\toprule[1.2pt]
\multirow{2}{*}{\textbf{Type}} & \multirow{2}{*}{\textbf{Benchmark}} & \multirow{2}{*}{\textbf{\makecell{Tradi-\\tional\\ \cite{b12} \cite{b13}}}} &
  \multicolumn{2}{c}{\textbf{\makecell{Baseline \\ (Pass Rate \%)}}} &
  \multicolumn{2}{c|}{\textbf{\makecell{Proposed \\ (Pass Rate \%)}}} &  
\multirow{2}{*}{\textbf{Type}} & \multirow{2}{*}
{\textbf{Benchmark}} & \multirow{2}{*}{\textbf{\makecell{Tradi-\\tional\\\cite{b12} \cite{b13}}}} &

\multicolumn{2}{c}{\textbf{\makecell{Baseline \\ (Pass Rate \%)}}} &
\multicolumn{2}{c}{\textbf{\makecell{Proposed \\ (Pass Rate \%)}}}
\\
\cmidrule(lr){4-5} \cmidrule(lr){6-7} \cmidrule(lr){11-12} \cmidrule(lr){13-14}
& & & \textbf{Compi.} & \textbf{Simu.} & \textbf{Compi.} & \textbf{Simu.} & & & & \textbf{Compi.} & \textbf{Simu.} & \textbf{Compi.} & \textbf{Simu.} \\

\midrule
\multirow{3}{*}{\makecell[l]{T1:\\Pointer}} & 1: Double Pointer &$\times / \times$ & 60 & 53.33 & 73.33 & 73.33 
& \multirow{3}{*}{\makecell[l]{T5:\\Boolean\\Operation}} & 13: Support Vector Mac. &$\times / \times$ & 46.67 & 46.67 & 93.33 & 93.33 \\
& 2: Hash Table &$\checkmark / \checkmark$ & 93.33 & 93.33 & 100 & 100 & & 14: Fourier Transform  &$\times / \times$ & 40 & 33.33 & 80 & 80\\
& 3: Deep Neural Network &$\times / \times$ & 60 & 60 & 86.67 & 80 & & 15: Color Correction &$\times / \times$ & 46.67 & 46.67 & 66.67 & 66.67 \\

\midrule
\multirow{3}{*}{\makecell[l]{T2:\\Dynamic\\Array}} 
& 4: Linear Programming &$\checkmark / \times$ & 93.33 & 86.67 & 100 & 100 
& \multirow{3}{*}{\makecell[l]{T6:\\Incomplete\\Statement}} 
& 16: Fibonacci Sequence &$\checkmark / \checkmark$ & 86.67 & 80 & 100 & 100 \\
& 5: Binary Tree &$\times / \times$ & 66.67 & 60 & 80 & 80 & 
& 17: Cyclic Rotation &$\times / \times$ & 40 & 40 & 86.67 & 73.33 \\
& 6: K-Nearest Neighbor &$\times / \times$ & 60 & 46.67 & 73.33 & 66.67 & 
& 18: AES &$\times / \times$ & 33.33 & 26.67 & 60 & 60 \\

\midrule
\multirow{3}{*}{\makecell[l]{T3:\\Recursion}} 
& 7: Linked List &$\checkmark / \checkmark$ & 73.33 & 73.33 & 93.33 & 93.33 
& \multirow{3}{*}{\makecell[l]{T7:\\ Unsupport.\\Struct}} & 19: Data Stream &$\times / \times$ & 66.67 & 53 & 73.33 & 66.67 \\
& 8: Depth-First Search &$\times / \times$ & 46.67 & 40 & 80 & 66.67 & & 20: Longest Increa. Path &$\times / \times$ & 60 & 60 & 86.67 & 86.67 \\
& 9: Breadth-First Search &$\times / \times$ & 60 & 60 & 73.33 & 73.33 & & 21: Max Points on Line &$\times / \times$ & 46.67 & 40 & 80 & 73.33 \\
\midrule
\multirow{3}{*}{\makecell[l]{T4:\\Bit Width}} & 10: Edge Detection &$\times / \times$ & 53.33 & 33.33 & 73.33 & 60 & \multirow{3}{*}{\makecell[l]{T8:\\Exception-\\Handling}} 
& 22: QR Decomposition &$\times / \times$ & 66.67 & 60 & 80 & 80 \\
& 11: Greedy Algorithm &$\times / \times$ & 46.67 & 46.67 & 86.67 & 80 & & 23: Dump Filter  &$\times / \times$ & 33.33 & 33.33 & 66.67 & 66.67 \\
& 12: Bubble Sort &$\checkmark / \checkmark$ & 100 & 100 & 100 & 100 & & 24: Turbo Encoder &$\times / \times$ & 73.33 & 66.67 & 93.33 & 93.33 \\
\bottomrule[1.2pt]
\end{tabular}}
\begin{tablenotes}
\item[] \scriptsize The pass rate is calculated from the results of 15 rounds of HLS simulation.
\end{tablenotes}
\end{threeparttable}
\vspace{-0.5cm}
\end{table*}

\subsection{Bit Width Optimization}
In traditional HLS development, developers often define variables as basic data types in C/C++ programs, which often contain higher than the bit width required for the actual task, thus resulting in resource waste. In Fig.~\ref{fig:bop}(a), to reduce bit width for efficient hardware designs,
we propose an LLM-driven bit width optimization scheme, which can find the minimum and maximum value of a variable based on a range analysis. This technique sends the C/C++ code, corresponding task scenarios, and the descriptions of input data to the LLM as prompts. The LLM then automatically creates a C++-based bit width optimization program for this task. This created program identifies variable declaration nodes, finds the maximum and minimum values of variables, and automatically adjusts the default data type with the determined optimal bit widths. 

Fig.~\ref{fig:bop}(b) illustrates the profiling of the variable value ‘$m$' in a regular C++ program, which follows a normal distribution. The optimization program identifies that ‘$m$' has a minimum value of 0 and a maximum value of 481, so only 9-bit is needed instead of 32-bit. The framework then implements integer variables with determined bit widths with the $ac\_uint$ or $ac\_int$ types provided by the HLS tool. The detailed demonstration of the bit width optimization scheme, including prompts and responses, is shown in Fig.~\ref{fig:stage3} in Appendix II.


\subsection{Joint LLM-Script Repair}
Although LLMs have been effective in automatic program repair, using LLMs can be costly, as users are charged by the number of input and output tokens. For example, the GPT-4 Turbo model, one of the most advanced LLMs, charges \textdollar{0.01} per 1K input tokens and \textdollar{0.03} per 1K output tokens~\cite{b16.2}. A cost-effective LLM-based repair technique should guide and steer the model toward the correct repair with the lowest possible cost, i.e., with the lowest possible number of tokens.

To minimize the repair cost of using LLMs, we first use traditional HLS repair scripts, which run efficiently on the local CPU to repair the HLS-incompatible errors in the regular C/C++ program. The LLM then conducts comprehensive repairs only on the program that has been processed by these scripts. By pre-repairing simple errors using the traditional scripts at the early stage, the repair cost of using LLMs can be effectively reduced.


\subsection{PPA Optimization}
Power, Performance, and Area (PPA) are critical metrics in hardware design \cite{b24.1}. To facilitate the automatic HLS optimization, the critical code segments with performance bottlenecks are identified using design analyzers in the HLS tool. Strategies such as adding or modifying pragmas are then applied to these critical code segments to optimize them.

The optimization strategies are identified by an HLS optimization strategy library, which is built by us according to the manual of the HLS tool \cite{b25}. As shown in Fig. \ref{fig:op}, the optimization strategy information $Opt\_L1$ consists of a brief concept, parameter descriptions, and optimization examples. The most suitable optimization strategies will be matched via a similarity search, such as a sentence transformer \cite{b26}, and then integrated into the prompts, utilizing the LLM's in-context learning capabilities to generate an optimized program. For example, loop unrolling (\textit{\#pragma hls\_unroll}) can be added by the LLM to transform a serial iteration of a loop into a parallel execution, which helps reduce circuit latency. 
This pragma needs to be set to yes or a value less than or equal to the maximum number of loop iterations. 
In addition, when memory interfaces become limiting factors, the LLM can optimize memory read and write operations by adding dual-port capabilities or increasing cache mechanisms in the HLS-C program. Another example of the area optimization for the hardware design, including prompts and responses, is demonstrated in Fig.~\ref{fig:stage5} in Appendix II.

After optimizing the HLS-C program, circuit optimization strategies such as retiming are further adopted to minimize the clock period and the number of sequential components of the synthesized circuit.

\section{Experimental Results}\label{sec:fourth}
We demonstrate the results of the proposed LLM-driven framework for HLS across 24 real-world applications in terms of repair pass rate, cost of using LLM, and optimized hardware performance. These benchmarks are from related work \cite{b12,b13,b14,b14.1} with HLS-incompatible errors. 

During the evaluation, GPT-4 Turbo Model was used as the LLM via OpenAI APIs \cite{b31}. All experiments were conducted with the Catapult HLS Tool on an Intel(R) Xeon(R) Silver 4314 2.40 GHz CPU. The logic circuits in implementing HLS-C programs were synthesized with Synopsys Design Compiler using the Nangate 45nm open-cell library. To demonstrate the repair capability of the proposed framework, each experiment for a particular application was repeated $n$ instances ($n$ = 15). In each instance, the LLM was queried five times to repair the program based on the errors reported by the HLS tool. We calculate the expectation repair pass rate as $\text{Pass Rate (\%)}= m/n$, where $m$ is the number of successfully repaired instances and $n$ is the number of all the instances.

Table II compares the repair pass rates of the proposed framework with traditional scripts~\cite{b12} \cite{b13} and the baseline to use GPT-4 Turbo directly to repair the benchmarks. The first and second columns of the left and right subtables show the incompatible error types (T1-T8) and their corresponding benchmarks (1-24). The third column shows the compilation and simulation pass rates using traditional scripts to repair the programs, with only a few examples passing the simulation test. The fourth, fifth, sixth, and seventh columns represent the compilation and simulation pass rates of the GPT-4 Turbo baseline and our proposed framework, respectively. According to these columns, the proposed LLM-driven framework can improve the pass rate effectively compared with the repair with traditional scripts and the direct application of the LLM.

\begin{figure*}[!t]
\vspace{-0.1cm}
    \centering
 \includegraphics[width=1\linewidth]{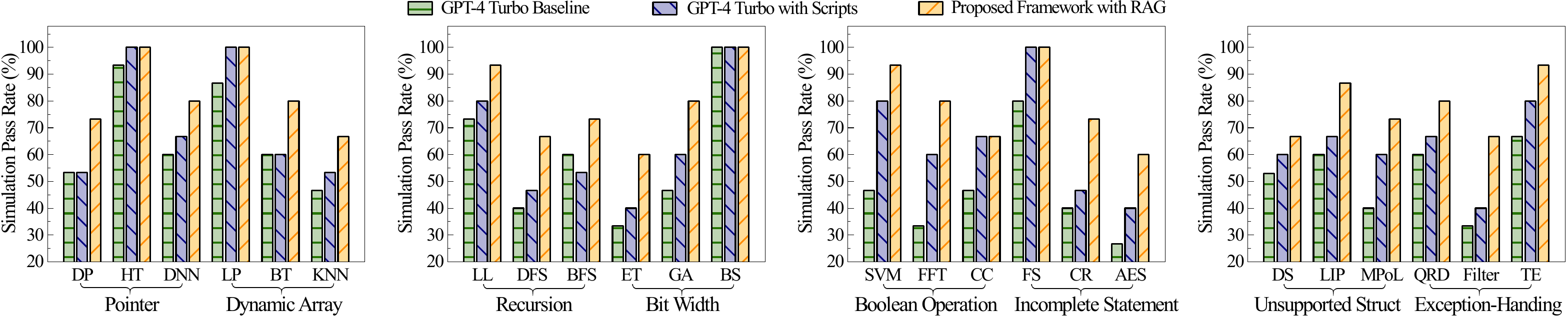}
        \vspace{-0.5cm}
        \caption{Comparison of HLS simulation pass rate of the proposed method with GPT-4 Turbo baseline and GPT-4 Turbo with scripts on 24 real-world applications, shown on the x-axes of the figures.}
        \label{fig:pass}
        \vspace{-0.5cm}
\end{figure*}

\begin{figure}[]
\centering	\includegraphics[width=1\linewidth]{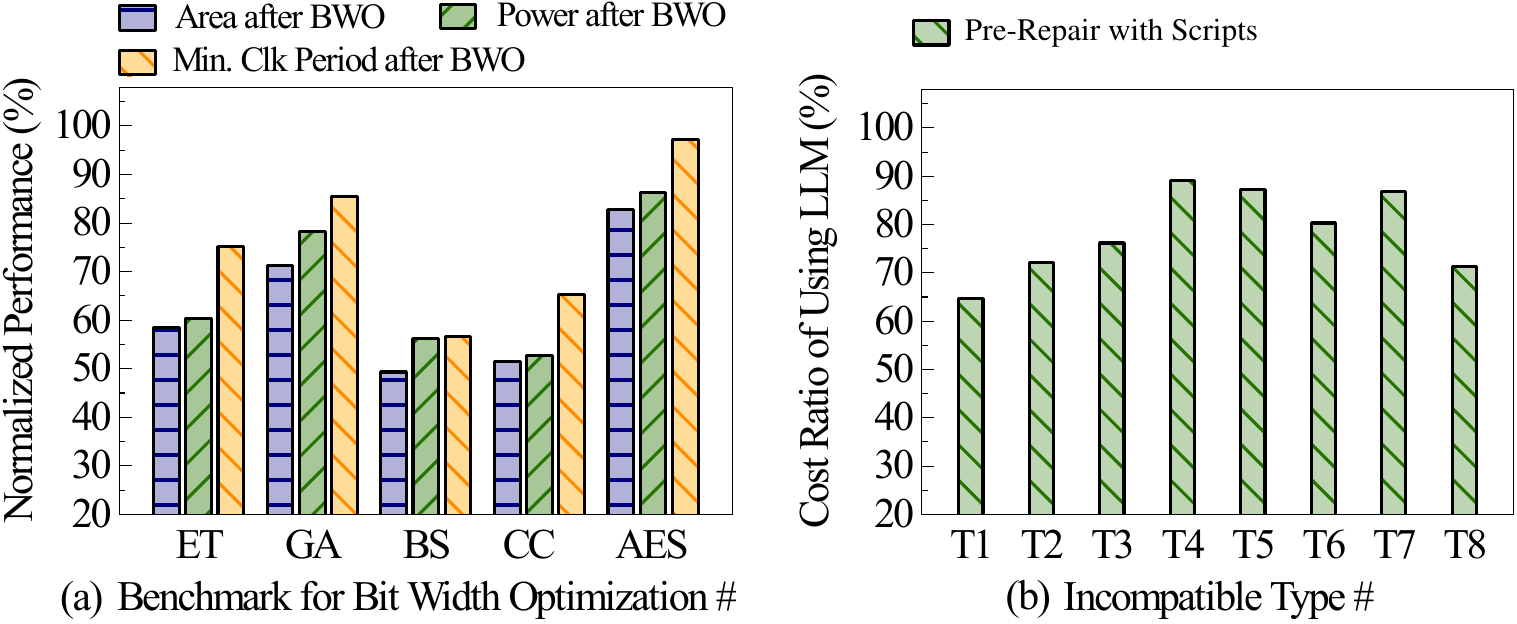}
	\caption{(a) Ratios of area, power consumption, and minimum clock period using the proposed bit width optimization compared with 32-bit \textit{int} implementation; (b) Cost ratios of the proposed joint LLM-script repair compared with using LLM-driven repair alone.}
	\label{fig:costbit}
\end{figure}

Fig.~\ref{fig:pass} compares the simulation pass rate of the proposed framework with those using GPT-4 Turbo directly and GPT joint scripts (ablation experiments without RAG) for program repair, respectively. According to this comparison, the proposed LLM-driven framework outperforms the baseline by achieving an average 23.33\% and 13.89\% increase in repair pass rate, respectively.

To demonstrate the advantages of the bit width optimization (BWO) scheme, we compare the proposed scheme with the default 32-bit \textit{int} type in terms of area, power, and minimum clock period. We use the actual scenarios and the descriptions of input data from the dataset as prompts to create a static bit width optimization program via the LLM. Fig~\ref{fig:costbit}(a) shows the performance ratios of the logic circuit after using bit width optimization. In this comparison, the proposed scheme automatically identified the optimized bit widths for integers, leading to an average reduction of 36.57\%, 33.03\%, and 29.08\% in area, power, and minimum clock period, respectively.

To verify the cost reduction of using LLM, we compared the cost of the proposed joint LLM-Script repair mechanism with that of using only LLM. Among all successful repair programs, we counted the average numbers of input and output tokens associated with the LLM separately. The total costs of using LLMs are calculated from these tokens and their corresponding costs.
The normalized average costs to repair each error type~\cite{b31} are shown in Fig.~\ref{fig:costbit}(b). According to this comparison, pre-repairing simple errors with scripts locally can significantly reduce the overall repair costs of only using the LLM by 21.56\% on average.

\begin{figure}[t]
\centering	\includegraphics[width=1\linewidth]{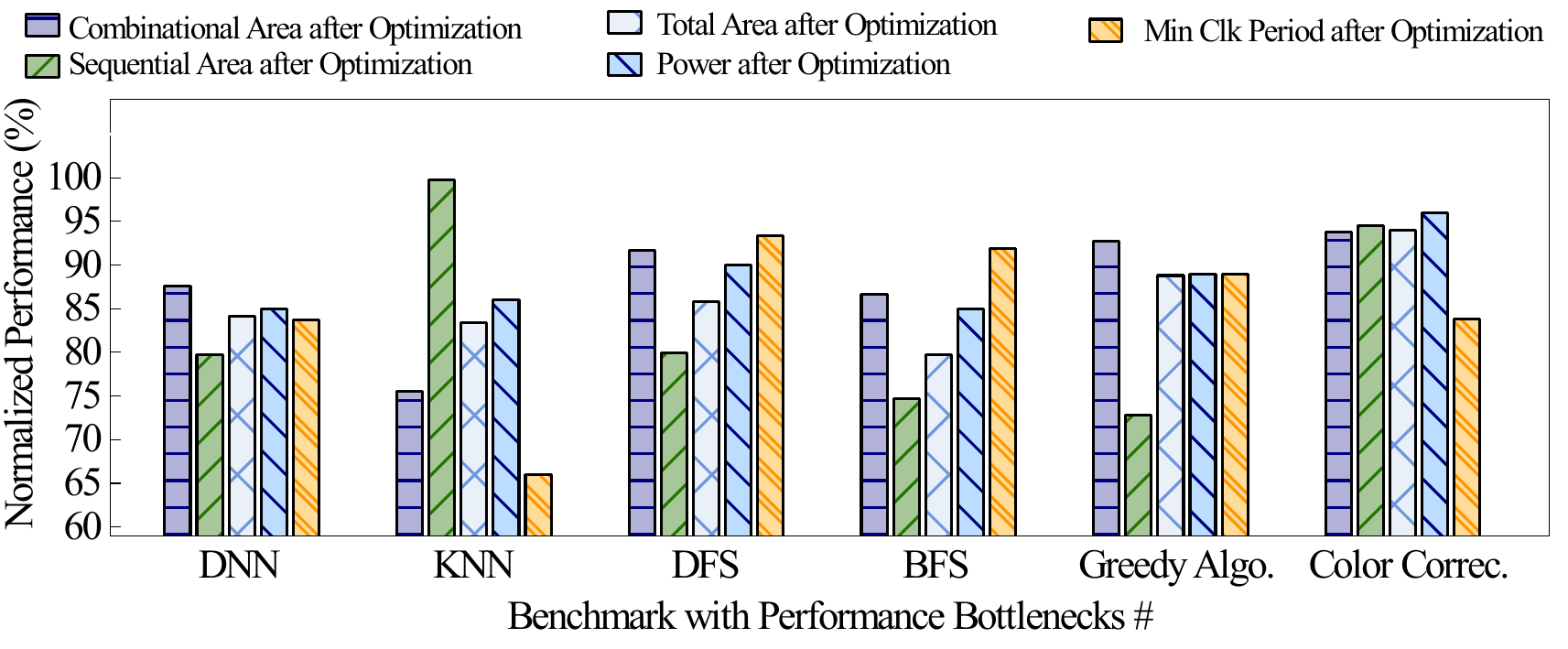}
	\caption{Ratios of area, power, and minimum clock period after applying LLM-driven optimization.}
	\label{fig:opre}
\end{figure}

To demonstrate the effectiveness of the proposed LLM-driven PPA optimization strategy on logic circuits, we compared the area, power, and minimum clock period of six benchmarks with performance bottlenecks before and after using such optimization. As shown in Fig.~\ref{fig:opre}, the proposed optimization strategy further achieves an average reduction of 14.02\%, 11.50\%, and 17.94\% in area, power, and minimum clock period, respectively, resulting in more efficient designs.

\section{Conclusion}\label{sec:fifth}
In this paper, we have proposed an LLM-driven program repair framework to solve the incompatible issues of regular C/C++ programs in HLS. A retrieval-augmented generation paradigm is introduced to guide the LLM toward correct repair. An LLM-driven bit width optimization scheme is then applied to identify the optimized bit widths for variables. A joint LLM-script repair mechanism is further used to reduce the cost of using the LLM. The critical code segments are extracted and optimized to generate more efficient HLS designs. Experimental results demonstrate that the proposed LLM-driven repair framework can achieve much higher repair pass rates on 24 real-world applications compared with the traditional scripts and the direct applications of the LLM.

\section*{Appendix I: An example of the proposed LLM-driven program repair framework for HLS}

\begin{figure}[h]
\vspace{-0.2cm}
\centering	
\includegraphics[width=0.99\linewidth]{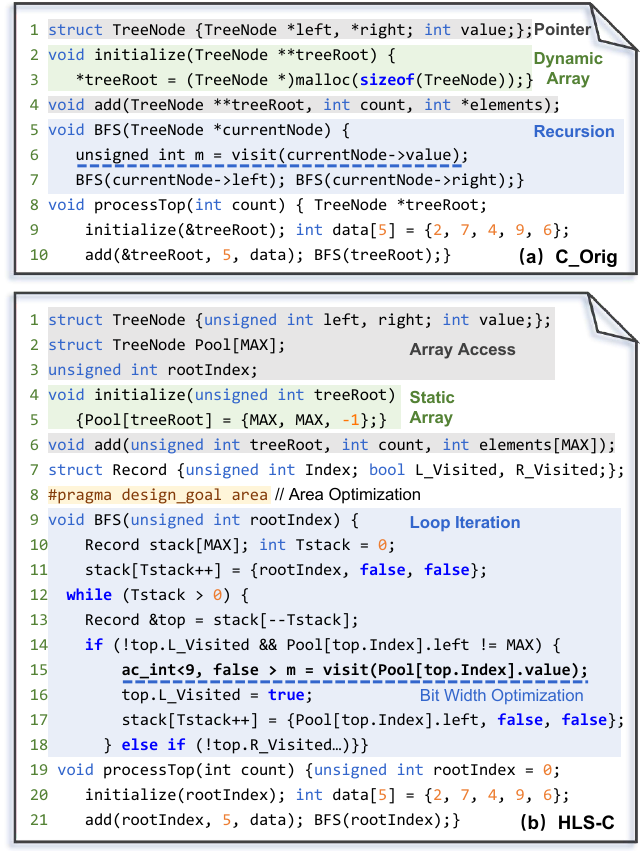}
\vspace{-0.5cm}
\caption{~Working example (Breadth-First Search) of the proposed LLM-driven repair framework. (a) Original C++ program using pointers, dynamic arrays, and recursion; (b) Repaired HLS-C program.}
\label{fig:bfs}
\vspace{-0.1cm}
\end{figure}

Consider a part of the Breadth-First Search program (in a total of 139 lines in the original code) shown in Fig.~\ref{fig:bfs}. This program involves creating a binary tree, initializing the tree structure, adding nodes, and then performing a Breadth-First Search (BFS) to visit and process each node. 

In Section~\ref{sec:fourth}, we conduct a comprehensive evaluation of all eight types of errors. Here, to clearly illustrate the repair process, we use this example containing three typical incompatible errors, i.e., ‘$pointer$', ‘$dynamic$ $array$' and ‘$recursion$'. The detailed repair process is as follows:

\textit{1) Preprocessing:} The original C++ program, $C\_Orig$, is compiled using the Catapult HLS tool, and the first compilation reports two incompatible errors: ‘$pointer$' and ‘$dynamic$ $array$'. Subsequently, both the common HLS-incompatible error types along with the complete program are fed into the LLM to detect other potential errors, during which the LLM identifies an additional incompatible error, ‘$recursion$'. Traditional scripts are used to repair the above three errors, but the compilation fails again. The process then moves to the stage of matching error correction templates.

\textit{2) Repair with RAG:} In this stage, a repair library is built with human guidance and HLS official documentation, which is used to match detected errors with their corresponding correction templates. The matched templates, $T\_Ref$, are then used to enhance the quality of prompts provided to the LLM, guiding the LLM to repair the $C\_Orig$, thereby generating a more accurate HLS-C program. As shown in Fig.~\ref{fig:bfs}, LLM implements three main repairs on the original program $C\_Orig$: 

\ding{172} Modify Pointer Access to Array Access: In $C\_Orig$, tree nodes are managed through pointers directly manipulating memory locations (Line 1 in Fig.~\ref{fig:bfs}(a): \texttt{TreeNode *left, *right}). The HLS-C version modified by the LLM replaces these pointer-based node accesses with array indices (Line 1-3 in Fig.~\ref{fig:bfs}(b)). This modification uses a pre-allocated array (Pool) of TreeNode structures, where each node can be accessed via its index rather than pointers.

\ding{173} Rewrite Dynamic Array into Static Array: In $C\_Orig$, \texttt{malloc} is used to create a new TreeNode with dynamic memory allocation (Line 2 and 3 in Fig.~\ref{fig:bfs}(a): \texttt{malloc(sizeof())}), which is not supported by HLS tools because the synthesized circuit cannot manage data structures with unbounded size. The LLM manages all tree nodes through a predefined array Pool (Line 4 and 5 in Fig.~\ref{fig:bfs}(b): \texttt{Pool[treeRoot] = \{MAX, MAX, -1\};}), which avoids dynamic memory allocation.

\ding{174} Convert Recursion to Iteration: As shown in lines 5-7 of Fig.~\ref{fig:bfs}(a), the original function \texttt{BFS} employs recursion to call itself for processing the left and right child nodes of the current node. The LLM replaces the recursion by using an explicit stack data structure, as repaired in lines 9-18 of Fig.~\ref{fig:bfs}(b). This stack simulates the automatic management of the function call stack used in recursion. After these repairs, the HLS tool compiles this program again. If the compilation still fails, iterative repair continues until a successful compilation is achieved. 


\textit{3) Bit Width Optimization:} To optimize the design of hardware implementations, the repaired C++ program, along with task scenarios and corresponding descriptions of input data, are fed into the LLM, which automatically generates a C++-based bit width optimization script to find the maximum and minimum values of variables. As shown in line 6 of Fig.~\ref{fig:bfs}(a) and the corresponding line 15 in Fig.~\ref{fig:bfs}(b), if a variable ‘\texttt{m}' is declared as a 32-bit integer and the optimization script finds that its minimum value is 0 and its maximum is 481, then only 9-bit are needed instead of 32-bit. Then, the previously generated script assigns the optimal bit width by replacing \texttt{unsigned int m} with ‘\texttt{ac\_int$<$9, false$>$ m}' in $C\_Orig$.

\textit{4) Equivalence Verification:} After synthesizing the C++ code to obtain the corresponding hardware RTL model, C-RTL co-simulation is then performed to verify the correctness of the synthesized RTL design. 

\textit{5) PPA Optimization:} Key code segments with large area in the top function \texttt{BFS} are identified by the HLS tool. The proposed LLM-driven optimization strategy is then applied to reduce the area of design by adding the pragma such as ‘\texttt{\#pragma design\_goal area}' at the top of the function \texttt{BFS}. Retiming is further used to minimize the clock period the number of sequential components in implementing the synthesized circuit to achieve a more efficient PPA design. 

\section*{Appendix II: Examples of prompts and output responses of the LLM at different stages}

\begin{figure}[h]
\vspace{-0.2cm}
\centering	\includegraphics[width=1.02\linewidth]{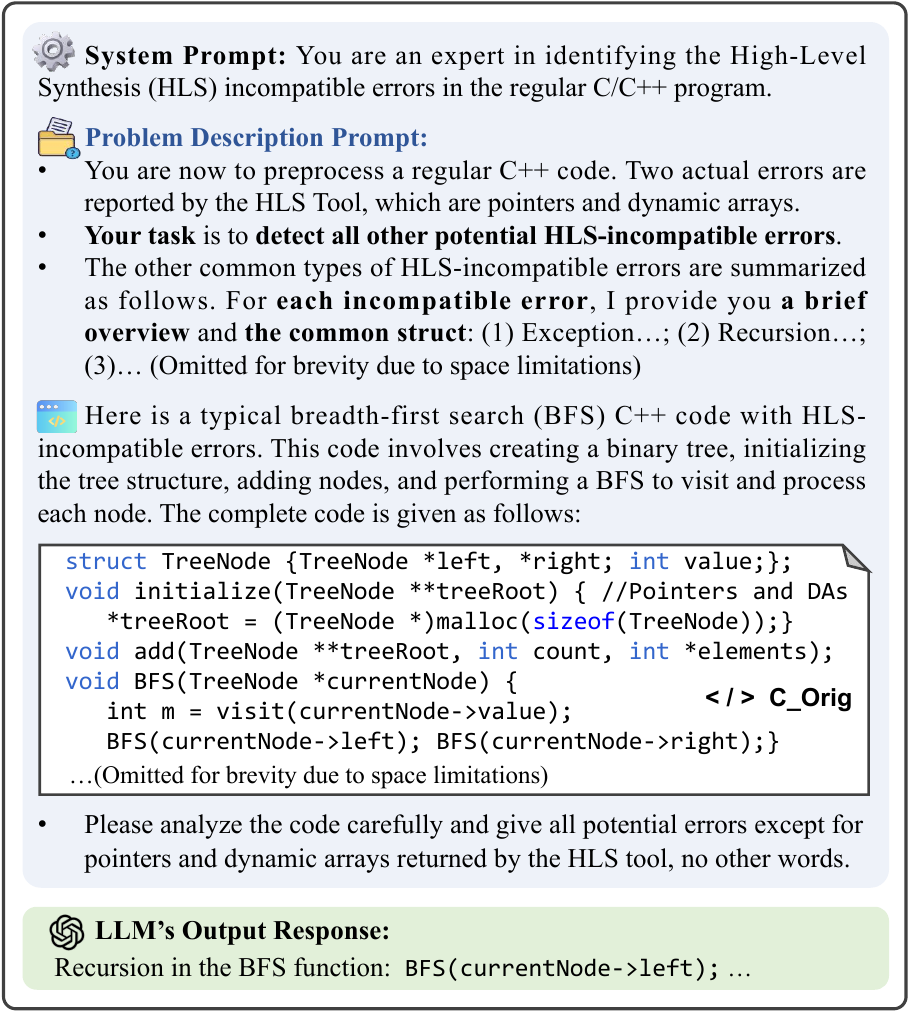}
\caption{Preprocessing: Identify potential HLS-incompatible errors in the regular C/C++ program via the LLM}
\label{fig:stage1}
\vspace{-0.3cm}
\end{figure}

\begin{figure}[h]
\centering	\includegraphics[width=1.02\linewidth]{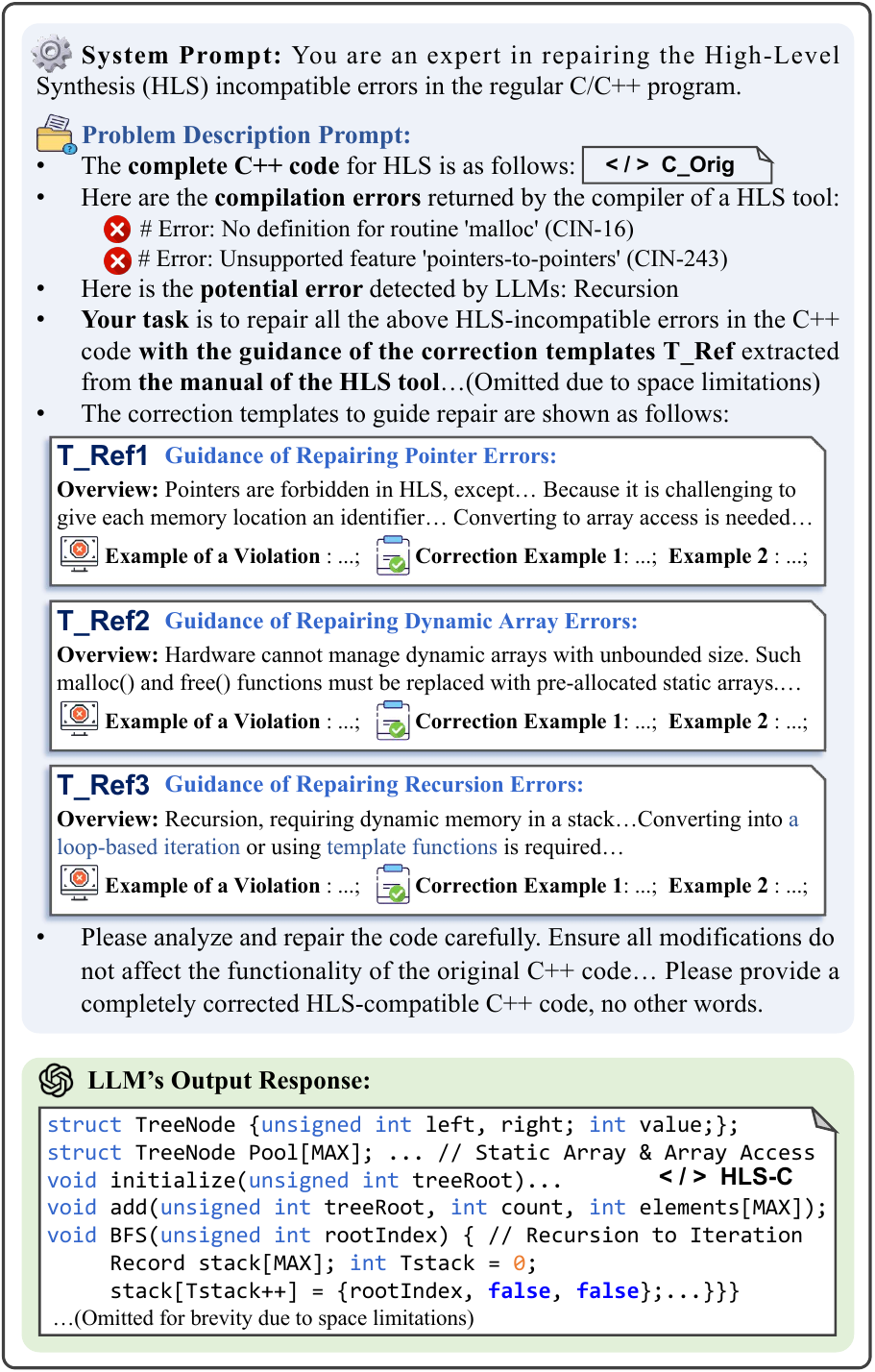}
\caption{Repair with RAG: LLM repairs the HLS-incompatible errors according to external guidance from the HLS tool manual.}
\label{fig:stage2}
\vspace{-0.1cm}
\end{figure}

\begin{figure}[h]
\centering	\includegraphics[width=1.02\linewidth]{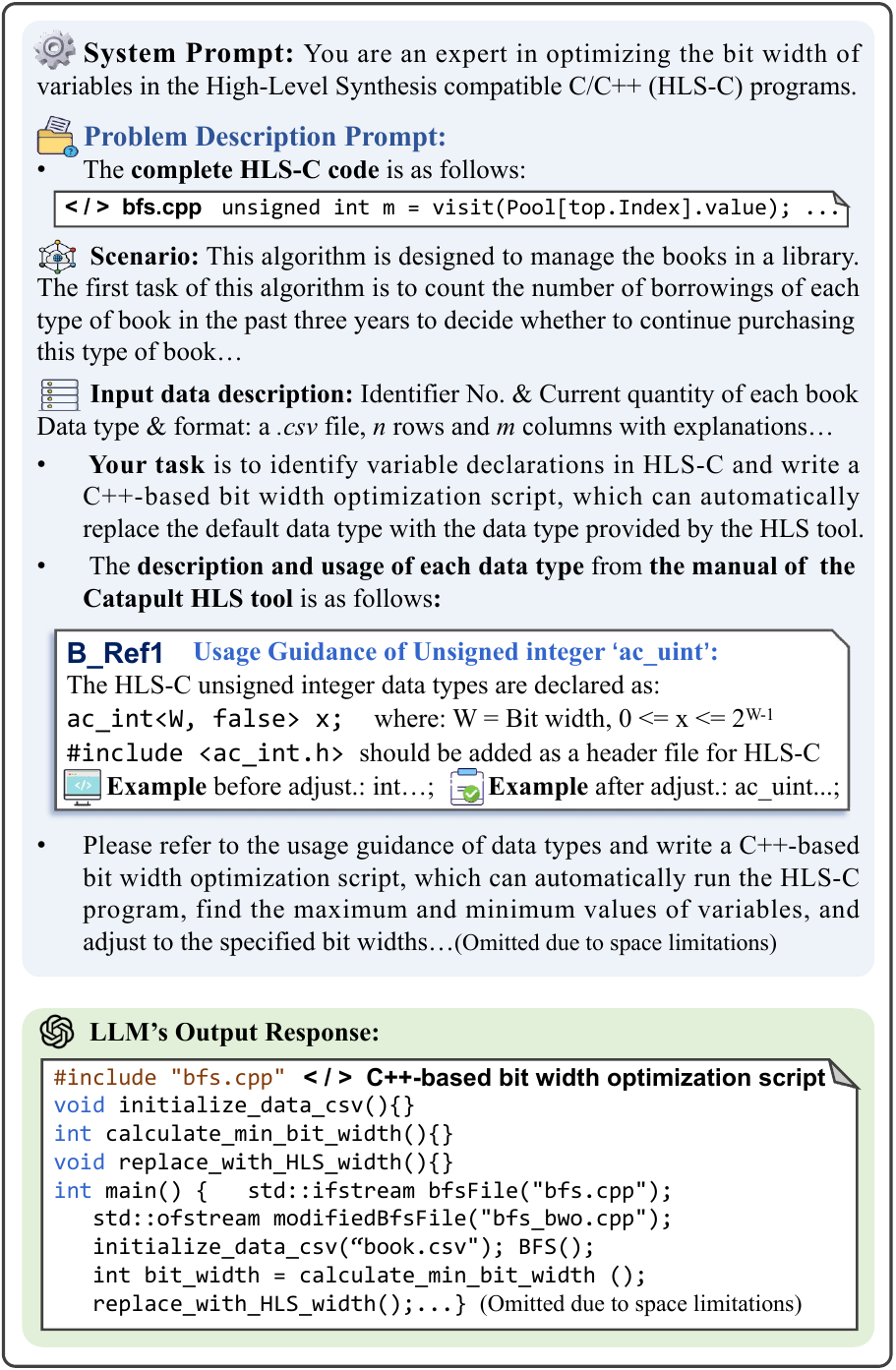}
\caption{Bit width optimization: A script is created via the LLM to adjust the default data type with the determined bit widths.}
\label{fig:stage3}
\end{figure}

\begin{figure}[h]
\centering	\includegraphics[width=1.02\linewidth]{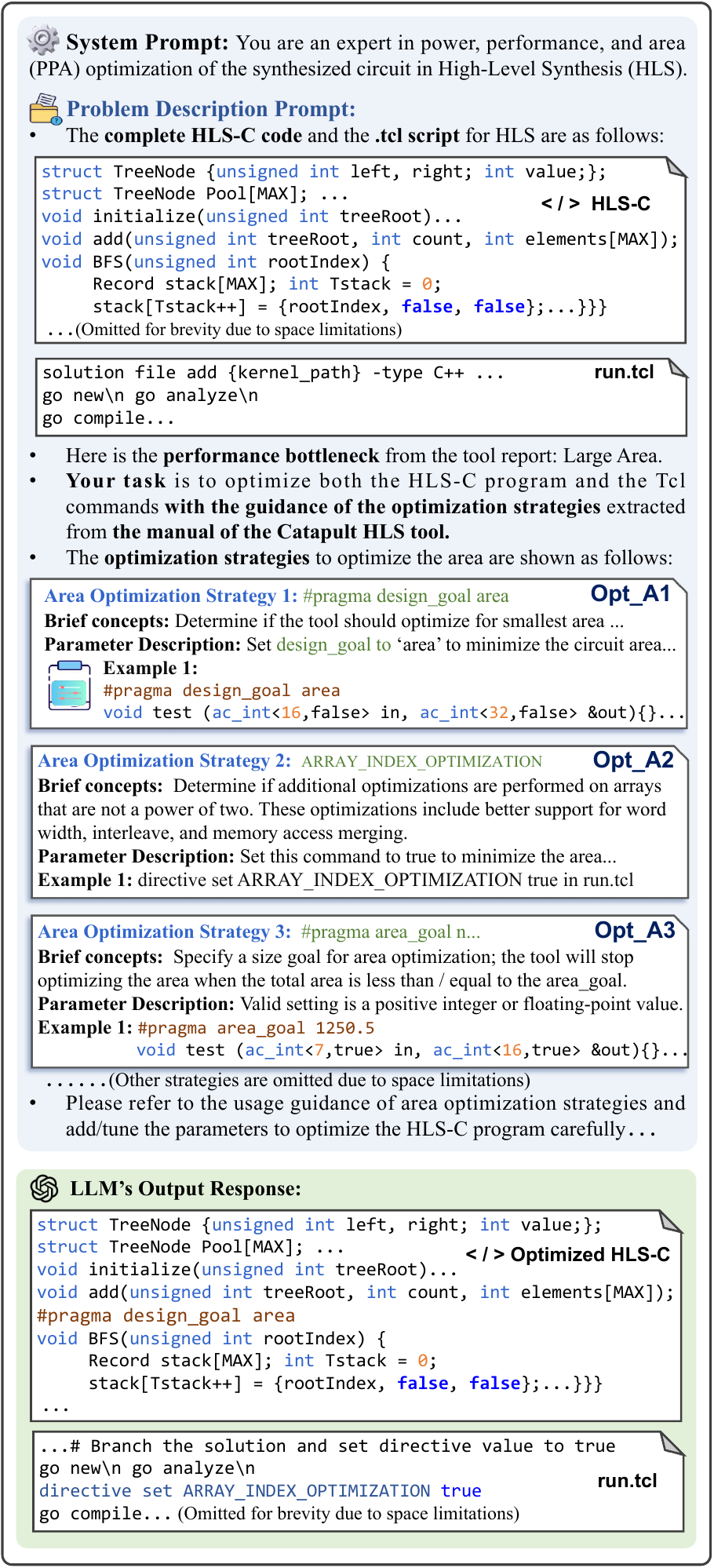}
\caption{PPA Optimization via the LLM.}
\label{fig:stage5}
\end{figure}


\begin{thebibliography}{9}
\bibitem{b10} Yunsheng Bai, Atefeh Sohrabizadeh, Zongyue Qin, Ziniu Hu, Yizhou Sun, Jason Cong, “Towards a Comprehensive Benchmark for High-Level Synthesis Targeted to FPGAs,” Advances in Neural Information Processing Systems (NeurIPS), 2023.
\bibitem{b10.1} Qi Sun, Tinghuan Chen, Siting Liu, Jianli Chen, Hao Yu, and Bei Yu, “Correlated Multi-objective Multi-fidelity Optimization for HLS Directives Design,” ACM Transactions on Design Automation of Electronic Systems (TODAES), 2022.
\bibitem{b10.2} Atefeh Sohrabizadeh, Cody Hao Yu, Min Gao, and Jason Cong, “AutoDSE: Enabling Software Programmers to Design Efficient FPGA Accelerators,” ACM Transactions on Design Automation of Electronic Systems (TODAES), 2022.
\bibitem{b11} Nadesh Ramanathan, George A. Constantinides and John Wickerson, “Precise pointer analysis in high-level
synthesis,” IEEE/ACM International Conference on Field-Programmable Logic and Applications (FPL), 2020.
\bibitem{b11.1} Jason Lau, Aishwarya Sivaraman, Qian Zhang, Muhammad Ali Gulzar, Jason Cong, Miryung Kim, “Refactoring for Heterogeneous Computing with FPGA," IEEE/ACM International Conference on Software Engineering (ICSE), 2020.
\bibitem{b12} David B. Thomas, “Synthesisable recursion for C++ HLS tools,” IEEE International Conference on Application-specific Systems, Architectures and Processors (ASAP), 2016.
\bibitem{b13} Zeping Xue, David B. Thomas, “SynADT: Dynamic data structures in high level synthesis,” IEEE Symposium on Field Programmable Custom Computing Machines (FCCM), 2016. 
\bibitem{b13.1} David B. Thomas, “Templatised Soft Floating-Point for High-Level Synthesis,” IEEE Symposium on Field Programmable Custom Computing Machines (FCCM), 2019. 
\bibitem{b14} Hsuan Hsiao, Jason H. Anderson, “Sensei: An area-reduction advisor for FPGA high-level synthesis,” IEEE/ACM Design, Automation \& Test in Europe Conference \& Exhibition (DATE), 2018.
\bibitem{b14.1} “Leetcode Problem Set,” Accessed: 2023. [Online]. Available: https://leetcode.com/problemset/.
\bibitem{b14.2} “Siemens EDA Catapult Synthesis User and Reference Manual,” Accessed: 2023. [Online]. Available: https://support.sw.siemens.com/.
\bibitem{b14.3} Kangwei Xu, Grace Li Zhang, Ulf Schlichtmann, Bing Li, “Logic Design of Neural Networks for High-Throughput and Low-Power Applications,” IEEE/ACM Asia and South Pacific Design Automation Conference (ASP-DAC), 2024.
\bibitem{b15} Yonggan Fu, Yongan Zhang, Zhongzhi Yu, Sixu Li, Zhifan Ye, Chaojian Li, Cheng Wan, Yingyan Lin, “GPT4AIGChip: Towards Next-Generation AI Accelerator Design Automation via Large Language Models,” IEEE/ACM International Conference on Computer Aided Design (ICCAD), 2023.
\bibitem{b16} Jason Blocklove, Siddharth Garg, Ramesh Karri, Hammond Pearce, “Chip-Chat: Challenges and Opportunities in Conversational Hardware Design,” IEEE Workshop on Machine Learning for CAD (MLCAD), 2023.
\bibitem{b16.1} Shailja Thakur, Jason Blocklove, Hammond Pearce, Benjamin Tan, Siddharth Garg, Ramesh Karri, “AutoChip: Automating HDL Generation Using LLM Feedback,” arXiv preprint: 2311.04887, 2023.
\bibitem{b16.2} “GPT-4 pricing for OpenAI API,” Accessed: 2024. [Online]. Available: https://openai.com/api/pricing/.
\bibitem{b16.3} Matthew Jin, Syed Shahriar, Michele Tufano, Xin Shi, Shuai Lu, Neel Sundaresan, and Alexey Svyatkovskiy, “InferFix: End-to-End Program Repair with LLMs,” ACM Joint European Software Engineering Conference and Symposium on the Foundations of Software Engineering, 2023.
\bibitem{b16.4} Yao Lu, Shang Liu, Qijun Zhang, Zhiyao Xie, "RTLLM: An Open-Source Benchmark for Design RTL Generation with Large Language Model,” IEEE/ACM Asia and South Pacific Design Automation Conference, 2024.
\bibitem{b17} Long Ouyang, Jeff Wu, Xu Jiang, Ryan Lowe et al., “Training language models to follow instructions with human feedback,” Advances in Neural Information Processing Systems (NeurIPS), 2022.
\bibitem{b18} Yun-Da Tsai, Mingjie Liu, Haoxing Ren, “Automatically Fixing RTL Syntax Errors with Large Language Model,” IEEE/ACM Design Automation Conference (DAC), 2024.
\bibitem{b19} Hammond Pearce, Benjamin Tan, Baleegh Ahmad, Ramesh Karri, Brendan Dolan-Gavitt, “Examining Zero-Shot Vulnerability Repair with Large Language Models,” IEEE Symposium on Security and Privacy, 2023
\bibitem{b19.1} Harshit Joshi, Jose Cambronero, Sumit Gulwani, Vu Le, Ivan Radicek, Gust Verbruggen, “Repair is nearly generation: multilingual program repair with LLMs,” Proceedings of the Thirty-Seventh AAAI Conference on Artificial Intelligence (AAAI), 2023.
\bibitem{b20} He Ye, Martin Monperrus, “ITER: Iterative Neural Repair for Multi-Location Patches,” IEEE/ACM International Conference on Software Engineering (ICSE), 2024.

\bibitem{b21} Chunqiu Steven Xia, Yuxiang Wei, Lingming Zhang, “Automated program repair in the era of large pre-trained language models,” IEEE/ACM International Conference on Software Engineering (ICSE), 2023.

\bibitem{b23} Nicholas V. Giamblanco, Jason H. Anderson, “A Dynamic Memory Allocation Library for High-Level Synthesis,” IEEE/ACM International Conference on Field-Programmable Logic and Applications (FPL), 2019.
\bibitem{b24.1} Tinghuan Chen, Grace Li Zhang, Bei Yu, Bing Li, Ulf Schlichtmann, “Machine Learning in Advanced IC Design: A Methodological Survey,” IEEE Design \& Test, 2023.
\bibitem{b25} “Siemens EDA Catapult High-Level Synthesis Tools,” Accessed: 2023. [Online]. Available: https://eda.sw.siemens.com/en-US/ic/catapult-high-level-synthesis/hls/.
\bibitem{b26} Nils Reimers, Iryna Gurevych, “Sentence-BERT: Sentence Embeddings using Siamese BERT-Networks,” ACL Empirical Methods in Natural Language Processing (EMNLP), 2019.
\bibitem{b31} “GPT-4 Turbo through OpenAI API,” Accessed: 2024. [Online]. Available: https://platform.openai.com/.
\end{thebibliography}
\end{document}